\documentstyle[12pt]{article}
\begin{document}
\def\be{\begin{equation}}
\def\ee{\end{equation}}
\def\bea{\begin{eqnarray}}
\def\ena{\end{eqnarray}}


\def\a{\alpha}
\def\b{\beta}
\def\c{\chi}
\def\d{\delta}
\def\e{\epsilon}           
\def\f{\phi}               
\def\g{\gamma}
\def\h{\eta}
\def\i{\iota}
\def\j{\psi}
\def\k{\kappa}                    
\def\l{\lambda}
\def\m{\mu}
\def\n{\nu}
\def\o{\omega}
\def\p{\pi}                
\def\q{\theta}                    
\def\r{\rho}                      
\def\s{\sigma}                    
\def\t{\tau}
\def\u{\upsilon}
\def\x{\xi}
\def\z{\zeta}
\def\D{\Delta}
\def\F{\Phi}
\def\G{\Gamma}
\def\J{\Psi}
\def\L{\Lambda}
\def\O{\Omega}
\def\P{\Pi}
\def\Q{\Theta}
\def\S{\Sigma}
\def\U{\Upsilon}
\def\X{\Xi}
\def\del{\partial}


\def\ca{{\cal A}}
\def\cb{{\cal B}}
\def\cc{{\cal C}}
\def\cd{{\cal D}}
\def\ce{{\cal E}}
\def\cf{{\cal F}}
\def\cg{{\cal G}}
\def\ch{{\cal H}}
\def\ci{{\cal I}}
\def\cj{{\cal J}}
\def\ck{{\cal K}}
\def\cl{{\cal L}}
\def\cm{{\cal M}}
\def\cn{{\cal N}}
\def\co{{\cal O}}
\def\cp{{\cal P}}
\def\cq{{\cal Q}}
\def\car{{\cal R}}
\def\cs{{\cal S}}
\def\ct{{\cal T}}
\def\cu{{\cal U}}
\def\cv{{\cal V}}
\def\cw{{\cal W}}
\def\cx{{\cal X}}
\def\cy{{\cal Y}}
\def\cz{{\cal Z}}

\def\Bf#1{\mbox{\boldmath $#1$}}       
\def\Sf#1{\hbox{\sf #1}}               
\def\TT#1{\hbox{#1}}                   


\def\bop#1{\setbox0=3D\hbox{$#1M$}\mkern1.5mu
	\vbox{\hrule height0pt depth.04\ht0
	\hbox{\vrule width.04\ht0 height.9\ht0 \kern.9\ht0
	\vrule width.04\ht0}\hrule height.04\ht0}\mkern1.5mu}
\def\Box{{\mathpalette\bop{}}}                        
\def\pa{\partial}                              
\def\de{\nabla}                                       
\def\dell{\bigtriangledown} 
\def\su{\sum}                                         
\def\pr{\prod}                                        
\def\iff{\leftrightarrow}                      
\def\conj{{\hbox{\large *}}} 
\def\lconj{{\hbox{\footnotesize *}}}          
\def\dg{\sp\dagger} 
\def\ddg{\sp\ddagger} 


\def\sp#1{{}^{#1}}                             
\def\sb#1{{}_{#1}}                             
\def\oldsl#1{\rlap/#1}                 
\def\sl#1{\rlap{\hbox{$\mskip 1 mu /$}}#1}
\def\Sl#1{\rlap{\hbox{$\mskip 3 mu /$}}#1}     
\def\SL#1{\rlap{\hbox{$\mskip 4.5 mu /$}}#1}   
\def\Tilde#1{\widetilde{#1}}                   
\def\Hat#1{\widehat{#1}}                       
\def\Bar#1{\overline{#1}}                      
\def\bra#1{\Big\langle #1\Big|}                       
\def\ket#1{\Big| #1\Big\rangle}                       
\def\VEV#1{\Big\langle #1\Big\rangle}                 
\def\abs#1{\Big| #1\Big|}                      
\def\sbra#1{\left\langle #1\right|}            
\def\sket#1{\left| #1\right\rangle}            
\def\sVEV#1{\left\langle #1\right\rangle}      
\def\sabs#1{\left| #1\right|}                  
\def\leftrightarrowfill{$\mathsurround=3D0pt \mathord\leftarrow \mkern-6mu
       \cleaders\hbox{$\mkern-2mu \mathord- \mkern-2mu$}\hfill
       \mkern-6mu \mathord\rightarrow$}
\def\dvec#1{\vbox{\ialign{##\crcr
       \leftrightarrowfill\crcr\noalign{\kern-1pt\nointerlineskip}
       $\hfil\displaystyle{#1}\hfil$\crcr}}}          
\def\hook#1{{\vrule height#1pt width0.4pt depth0pt}}
\def\leftrighthookfill#1{$\mathsurround=3D0pt \mathord\hook#1
       \hrulefill\mathord\hook#1$}
\def\underhook#1{\vtop{\ialign{##\crcr                 
       $\hfil\displaystyle{#1}\hfil$\crcr
       \noalign{\kern-1pt\nointerlineskip\vskip2pt}
       \leftrighthookfill5\crcr}}}
\def\smallunderhook#1{\vtop{\ialign{##\crcr      
       $\hfil\scriptstyle{#1}\hfil$\crcr
       \noalign{\kern-1pt\nointerlineskip\vskip2pt}
       \leftrighthookfill3\crcr}}}
\def\der#1{{\pa \over \pa {#1}}}               
\def\fder#1{{\d \over \d {#1}}} 


\def\frac#1#2{{\textstyle{#1\over\vphantom2\smash{\raise.20ex
       \hbox{$\scriptstyle{#2}$}}}}}                  
\def\ha{\frac12}                               
\def\sfrac#1#2{{\vphantom1\smash{\lower.5ex\hbox{\small$#1$}}\over
       \vphantom1\smash{\raise.4ex\hbox{\small$#2$}}}} 
\def\bfrac#1#2{{\vphantom1\smash{\lower.5ex\hbox{$#1$}}\over
       \vphantom1\smash{\raise.3ex\hbox{$#2$}}}}      
\def\afrac#1#2{{\vphantom1\smash{\lower.5ex\hbox{$#1$}}\over#2}}  
\def\dder#1#2{{\pa #1\over\pa #2}}        
\def\secder#1#2#3{{\pa\sp 2 #1\over\pa #2 \pa #3}}          
\def\fdder#1#2{{\d #1\over\d #2}}         
\def\on#1#2{{\buildrel{\mkern2.5mu#1\mkern-2.5mu}\over{#2}}}
\def\On#1#2{\mathop{\null#2}\limits^{\mkern2.5mu#1\mkern-2.5mu}}
\def\under#1#2{\mathop{\null#2}\limits_{#1}}          
\def\bvec#1{\on\leftarrow{#1}}                 
\def\oover#1{\on\circ{#1}}                            
\def\dt#1{\on{\hbox{\LARGE .}}{#1}}                   
\def\dtt#1{\on\bullet{#1}}                      
\def\ddt#1{\on{\hbox{\LARGE .\kern-2pt.}}#1}             
\def\tdt#1{\on{\hbox{\LARGE .\kern-2pt.\kern-2pt.}}#1}   


\def\NP{Nucl. Phys. B}
\def\PL{Phys. Lett. }
\def\PR{Phys. Rev. Lett. }
\def\PRD{Phys. Rev. D}
\def\Ref#1{$\sp{#1)}$}

\def\eq{\begin{equation}}
\def\eqe{\end{equation}}
\def\eqa{\begin{eqnarray}}
\def\eqae{\end{eqnarray}}

\def\be{\begin{equation}}
\def\ee{\end{equation}}
\def\bea{\begin{eqnarray}}
\def\ena{\end{eqnarray}}

\def\to{\rightarrow}

\def\half{{1\over 2}}
\baselineskip=12pt
\hfill {${{\scriptstyle\rm
(ITP-SB-94-05)}\atop{\scriptstyle\rm(BONN-HE-94-06)}}$}

\vspace{.30in}
{\Large
\centerline{\bf Correlation Function of the Spin-1/2}
\centerline{\bf XXX Antiferromagnet}}

\vspace{.40in}

\centerline{\it Vladimir E. Korepin \footnote{e-mail:
korepin@max.physics.sunysb.edu}}
\centerline{Institute for Theoretical Physics}
\centerline{State University of New York at Stony Brook}
\centerline{Stony Brook, NY 11974-3840, USA}

\vspace{.20in}

\centerline{\it Anatoli G. Izergin \footnote{e-mail: izergin@lomi.spb.su}}
\centerline{Department of Mathematical Institute of Sciences}
\centerline{Academy of Sciences of Russia}
\centerline{St. Petersburg,  POMI}
\centerline{SU-199106, Fontanka 27, Russia}

\vspace{.20in}

\centerline{\it Fabian H.L. Essler \footnote{e-mail:
fabman@avzw01.physik.uni-bonn.de}}
\centerline{Physikalisches Institut der Universit\"at Bonn}
\centerline{Nussallee 12, 53115 Bonn, Germany}

\vspace{.20in}

\centerline{\it Denis B. Uglov
\footnote{e-mail:denis@max.physics.sunysb.edu}}
\centerline{Institute for Theoretical Physics}
\centerline{State University of New York at Stony Brook}
\centerline{Stony Brook, NY 11794-3840, USA}

\vspace{.30in}

\centerline{\bf Abstract}

\vspace{.5cm}

\noindent We consider a special correlation function in the isotropic
spin-$\half$ Heisenberg antiferromagnet. It is the
probability of finding a ferromagnetic string of (adjacent) spins
in the antiferromagnetic ground state. We give two different
representations for this correlation function. Both of them are exact
at any distance, but one becomes more effective for the description of
long distance behaviour, the other for the description of short
distance behaviour.

\baselineskip=16pt
\section{Introduction}
We consider the spin-$1\over 2$ Heisenberg $XXX$ antiferromagnet in a
magnetic field. The hamiltonian of the model is given by
\eq
H = \sum_{j\e Z} \s_j^x \s_{j+1}^x + \s_j^y \s_{j+1}^y + \s_j^z \s_{j+1}^z
- h \s_j^z\ .
\eqe
Here $j$ labels the sites of the lattice and runs through all integers,
$\s_j$ are Pauli matrices, and $h$ is the magnetic field directed
along the $z$ direction in spin space. Eigenfunctions of the model
were constructed by H. Bethe \cite{Bethe}. The antiferromagnetic
ground state $|AFM\rangle$ at zero temperature was determined by
Hulth\'en in \cite{hul}. The ground state as well as excitations are
described by linear integral equations. The integral equation for the
energy $\e (\l)$ as a function of the spectral parameter $\l$ of the
elementary excitation (a spin-$\half$ kink)\cite{ft} reads
\eq
\e (\l) + \frac{1}{2\p} \int^\L_{-\L} K (\l, \m) \e (\m ) d \m = 2h -
{2\over {1\over 4} + \l^2}
\eqe
Here $K (\l, \m ) = \frac{2}{1+ (\l - \m)^2}$.  The energy should be
equal to zero at the Fermi edges $\pm \L$, {\em i.e.} $\e (\L ) = \e
(-\L) = 0$. This shows the $\L$ depends on $h$.  In the limit $h \to
0$ $\L$ tends to $\infty$ according to
\eq
\L = \frac{1}{2\p} \ln \frac{(2 \p)^3}{e h^2}
\eqe
On the other hand $\L \to 0$ as $h \to 4$, and $\L=0$ for $h\geq 4$.

In order to define the correlation function of interest we consider
the operators
\eq
P_j = \half (\s_j^z + 1) = \left( \begin{array}{cc} 1 & 0 \\ 0 & 0
\end{array} \right)\ ,
\eqe
which project on the state with spin up at site number $j$.

The so-called ``Emptiness-Formation Probability'' correlation function
is defined as
\eq
P (x) = \langle AFM|\; \prod^x_{j=1} P_j \;|AFM \rangle .
\eqe
\noindent The physical meaning of $P(x)$ is the probability of finding
$x$ (where $x$ is an integer) adjacent spins up in the
antiferromagnetic vacuum.

Below we give two different representations for this probability based
on two different methods of determining correlation functions. The
reason for studying $P(x)$ rather than correlators of local spins is
that $P(x)$ is the simplest correlation function from a computational
point of view in {\em both} approaches.

\section{The Dual Field Approach}

The first approach is based on the papers [4-7]. It is described in
detail in the book \cite{KIB}. There are three main steps:
\begin{itemize}
\item{} Starting from the solution of the model by means of
the Algebraic Bethe Ansatz, correlation functions are expressed as
determinants of Fredholm integral operators.

\item{} The determinants are described by means of
integro-differential equations (see Sect. XIV.7. of \cite{KIB}).

\item{} The large-distance asymptotics of the correlation
functions are obtained from the solution of a Riemann-Hilbert problem
[8-12].
\end{itemize}
Let us now consider step 1. In order to express $P(x)$ as a
determinant we need to make use of dual quantum fields $\varphi(\l)$,
which are linear combinations of canonical Bose fields
$a(\l)$ and $a^\dagger(\l)$ with standard commutation relations
\eqa
\bigg[ a (\l ), a^\dagger (\m ) \bigg] &=& \d (\l - \m) \nonumber\\
\bigg[ a (\l ) , a (\m ) \bigg] &=& 0 = \bigg[ a^\dagger (\l ),
a^\dagger (\m ) \bigg]\ .
\eqae
They are acting in the standard Fock space with reference state $\vert
0)$ (Fock vacuum) defined {\em via}
\eq
a (\l ) \vert 0)  = 0 \quad , \quad 0= (0 \vert a^\dagger (\m )\ .
\eqe
The dual fields $\varphi (\l )$ are given by the following
expressions:
\eq
\varphi (\l ) = a (\l ) - \int^\infty_{-\infty} d \n \ln \bigg[ 1 +
(\l-\n)^2 \bigg] a^\dagger(\n )\ .
\eqe
One should note that by construction the dual fields always commute
\eq
\bigg[ \varphi (\l ), \varphi (\m ) \bigg] = 0\ .
\eqe

In the book \cite{KIB} it is explained in detail for the case of the
Bose gas with delta-function interaction how to derive an expression
for $P(x)$ in terms of Fredholm determinants. $P(x)$ is found to be
a ratio of two determinants (see [8], formula (1.28) on page 246.).
For the $XXX$ model a similar expression is valid
\eq
P (x) = {(0\vert\det (1+\hat{V})\vert 0) \over \det (1+ {1 \over
2\p} \hat{K})} \ .
\eqe
Here ${\hat V}$ and ${\hat K}$ are integral operators acting on the
interval $[-\L, \L ]$, and $\vert 0)$ is the vacuum of the bosonic
Fock space defined in (7).  The kernel of ${\hat K}$ is the same as in
the  integral equation for the dressed energy of the elementary
excitation (2)
\eq
K (\l, \m ) = \frac{2}{1+ (\l - \m )^2}\ .
\eqe
The operator $\hat V$ also acts in the Fock space of the dual fields
and has a kernel
\eq
V (\l, \m)  =  \frac{1}{2\p} \bigg\{ {e (\l) e^{-1}(\m)\over (\l -
\m)(\l-\m + i)} + {e^{-1} (\l) e(\m) \over (\m - \l)(\m - \l + i)}
\bigg\}
\eqe
\eq
e (\l) = \bigg( {2\l + i \over 2 \l - i} \bigg)^{x/2} \exp(\varphi (\l
)/2) \ .
\eqe

Note that the expression for $V(\l, \m)$ involves only commuting
operators.

Formulas (10)-(13) are our main result and conclude step 1 of the
dual field approach to correlation functions.

A.R. Its is considering the Riemann-Hilbert problem generated by the
integral operator (12). The solution of the Riemann-Hilbert will give
an explicit formula for large distance asymptotics \cite{Its2}.

\section{The Vertex Operator Approach}

There exists another approach to the problem of determining
correlation functions at $h=0$. It was invented recently by the
RIMS group \cite{Jimbo}. In that paper correlation functions in the
$XXZ$ model were considered. We took the $XXX$ limit and obtained
the following representation for $P(x)$ in terms of $x$ multiple
integrals (recall that $x$  is integer).
\eqa
&& P (x) = \int_C {d \l_1 \over 2\p i \l_1} \int_C  {d \l_2 \over 2\p
i \l_2} \ldots \int_C
{d \l_x \over 2\p i\l_x} \bigg\{ \prod_{a=1}^{x} \bigg( 1 + {i \over
\l_a} \bigg)^{x-a}
\bigg( {\p \l_a \over \sinh(\p \l_a)} \bigg)^x\nonumber\\ \nonumber\\
&& \times \prod_{1 \leq j < k \leq x} \; {\sinh(\p (\l_k - \l_j)) \over \p
(\l_k - \l_j - i) } \bigg\}\ .
\eqae
Here each integral is performed along a contour $C$, which goes from
$-\infty$ to $+ \infty$ below the real axis.
Formula (14) (as well as formula (10)) is exact for all distances $x$,
but it becomes most efficient at small $x$.  Formula (14) is valid
at zero magnetic field $h=0$ only. The first few values are
$P(1)={1\over 2}$, $P(2)= {1\over 3}(1-\ln(2))$.

\noindent {\bf Acknowledgements :}  It is a pleasure to thank F.
Smirnov and L.D. Faddeev for useful discussions. After completing this
work we learned that A. Nakayashiki independently obtained a formula,
which is equivalent to (14).

\end{document}